\documentclass[prb,preprint]{revtex4}
\usepackage{graphicx}
\usepackage{amsmath}
\usepackage{amssymb}
\usepackage{bbm}

\newcommand{\bra}[1]{\left< #1 \right|}
\newcommand{\ket}[1]{\left|#1\right>}
\newcommand{\braket}[2]{\left< #1|#2\right>}

\begin{document}
 \title{Tomography scheme for two spin-1/2 qubits in a double quantum dot}
\author{Niklas Rohling and Guido Burkard}
\affiliation{Department of Physics, University of Konstanz, D-78457 Konstanz, Germany}

\begin{abstract}
We present a scheme for full quantum state tomography tailored for
two spin qubits in a double quantum dot.
A set of 15 quantum states allows to determine the density matrix
in this two-qubit space by projective measurement.
In this paper we choose a set gained from mutually unbiased bases.
We determine how those 15 projections can be represented by charge
measurements after a spin-to-charge conversion and the application
of quantum gates.
The quantum gates include exchange-based gates as well as rotations
by electron spin resonance (ESR).
We assume the experimental realization of ESR operations to be more
difficult than the exchange gate operation.
Therefore, it is an important result that the ESR gates are limited
by a $\pi/2$ rotation for one of the electron spins per measurement.

\end{abstract}
\maketitle

\section{Introduction}
Since electron spins in quantum dots were proposed\cite{LoDi1998} as qubits,
a lot of experimental progress has been made with a two-electron double
quantum dot as the basic cell of this kind of quantum computing:
Time evolution due to exchange interaction was demonstrated\cite{petta} as
well as electron spin resonance (ESR) of one electron spin.\cite{koppens}
Recently, Brunner and coworkers\cite{brunner} demonstrated both, exchange
interaction and ESR rotation in the same double quantum dot.
In principle this allows to perform arbitrary quantum gates in this system.
Ultimate control of a quantum system can be demonstrated in experiment
by state tomography, i.e., gaining full information of a
quantum state, in general described by a density matrix.
Since the density matrix is Hermitian with trace 1, it can can be described
by $N^2-1$ real parameters for a $N$-dimensional Hilbert space.
Therefore, from repeated projection of the unknown quantum state onto at least $N^2-1$
known states, the density matrix can be reconstructed from
experimental data.
A higher number of different measurements can increase the precision.\cite{deBurgh}
The experimental values of measurements are not perfect due to external noise and because
they are not performed infinitely often.
This may lead to unphysical results for the estimation of the density matrix by violating
the non-negativity condition.
By applying maximum likelihood\cite{hradil,james,smolin}
or Bayesian\cite{helstrom1969,tanaka_komaki,jones,derka_fine_mech_and_opt,derka_acta,buzek,schack,blum-kohout}
methods those results can be avoided.

State tomography is well established in nuclear magnetic resonance experiments
with qubits encoded by nuclear spins of trapped ions\cite{chuang,roos,haeffner_roos_blatt}
for up to eight qubits.\cite{haeffner_nature_2005}
It was also performed with superconducting qubits.\cite{steffen_prl,steffen_science}
For spin qubits, state tomography was done for qubits represented by the singlet and
one triplet level of two electron spins in a double quantum dot for the
single-qubit\cite{foletti2009} and the two-qubit case\cite{shulman}
as well as recently for an exchange-only qubit built by spin states of
three electrons in a triple quantum dot.\cite{medford}

Here, we present a measurement scheme for determining the unknown
mixed quantum state of a two-qubit system realized by two electron spins
in one double quantum dot.
A complete set of measurements for the two-qubit space is, e.g.,
given by James et al.,\cite{james} in that case referred to
the polarization of two photons.
We use another series of projections, which is based on a set of 
so-called mutually unbiased
bases\cite{ivanovic,wootters_fields,klappenecker_roetteler,durt}
and which includes entangled states in contrast to
Ref.~\onlinecite{james}.
In this issue our scheme also differs from the scheme used
in Ref.~\onlinecite{shulman} where the state of each of the two
qubits, both encoded in one two-electron double quantum dot, is
measured separately and the density matrix is determined using
correlation data.
The measurements we propose are projections on one-dimensional
subspaces of the four-dimensional Hilbert space, and thus 
joint measurements on the full two-qubit space.
An important result of our paper is that these measurements can be done
by applying ESR maximally for a $\pi/2$ rotation on one of the electron
spins per measurement.
We assume the ESR to be experimentally more costly than exchange gates
as in experiments the time for a complete ESR rotation was in the order of
100~ns\cite{koppens,brunner} and a complete rotation due to the exchange
interaction could be achieved within some ns.\cite{petta,brunner}
Furthermore, we assume the ESR gates to have a lower fidelity than
exchange gates.

In Sec.~\ref{sec:phys_sys} we describe the double-dot system with two
electrons and the possible manipulation of the quantum states.
Sec.~\ref{sec:general} provides information on the matrix space for
the traceless part of two-qubit density matrices and on properties of
a basis in this space represented by pure quantum states.
The main result of this paper, a scheme for constructing such a set of
quantum states using quantum gates and spin-to-charge conversion,
is given in Sec.~\ref{sec:quorum}.
In Sec.~\ref{sec:error}, we discuss how the influence of imperfection in
the quantum gates might be treated in experiment, before we conclude
in Sec.~\ref{sec:conclusion}.

\section{Physical system}
\label{sec:phys_sys}
We consider a double quantum dot with two electrons taking only the lowest (orbital) energy level in each dot into account.
Therefore, due to the antisymmetry of the wave function (Pauli principle),
we have a six-dimensional Hilbert space since we have one state with both electrons in the left dot,
one state with two electrons in the right dot, and a four-dimensional space with one electron per dot.
We will denote charge states with $n_l$ electron in the left and $n_r$
electrons in the right dot with $(n_l,n_r)$.
The Hamiltonian, taking into account the difference between the energy levels, $\varepsilon$,
the Coulomb penalty for both electrons in the same dot, $U$,
and the hopping matrix element between the dots, $t$,
reads
\begin{equation}
\label{eqn:Hamiltonian1}
 H =  \frac{\varepsilon}{2} (\hat n_1 - \hat n_2) + \frac{U}{2}\sum_{i=1,2} \hat n_i(\hat n_i - 1) + t\sum_{s=\uparrow,\downarrow}(\hat c_{2s}^\dagger\hat c_{1s} +\hat c_{1s}^\dagger\hat c_{2s})
\end{equation}
where the operators of the occupation number in the dots,
$\hat n_1$ and $\hat n_2$ are given by $\hat n_i = \sum_{s=\uparrow,\downarrow}\hat c^\dagger_{is}\hat c_{is}$.

We will consider the states with charge configuration (1,1) as logical Hilbert
space and regard the electron spin in each dot as a quantum bit.
The aim of this paper is to give a scheme for determining an unknown density matrix
of this two-qubit space.
In the limit $|t|\ll|U\pm\varepsilon|$ the (1,1) states are approximately
decoupled from the states with charge configuration (2,0) and (0,2).
For this case the effective Hamiltonian for (1,1) states can be obtained by
applying a Schrieffer-Wolff transformation, \cite{burkard_imamoglu}
\begin{equation}
\label{eqn:exch}
 H_{\rm eff} = J \frac{\boldsymbol{\sigma}_1\cdot\boldsymbol{\sigma}_2 - 1}{4}=-JP_S
\end{equation}
with $J=4t^2U/(U^2-\varepsilon^2)$, and where $\boldsymbol{\sigma}_i$ is the vector
of Pauli matrices for the electron spin in dot $i=1,2$.
Note that $H_{\rm eff}$ is given by ${-}J$ times the projection operator $P_S$ on
the spin singlet
$\ket{S} = \frac{\ket{\uparrow\downarrow}-\ket{\downarrow\uparrow}}{\sqrt{2}}$
because only an antisymmetric state can couple to the (0,2) and to the (2,0) space.
The strength of this exchange coupling, $J$, can be tuned in experiment by
varying $\varepsilon$ or $t$.

We now consider in addition to the exchange interaction above a magnetic
field yielding the Zeeman Hamiltonian
\begin{equation}
 H_B = \textbf{h}_1\cdot\boldsymbol{\sigma}_1 + \textbf{h}_2\cdot\boldsymbol{\sigma}_2.
\end{equation}
\begin{figure}
\includegraphics[height=15cm]{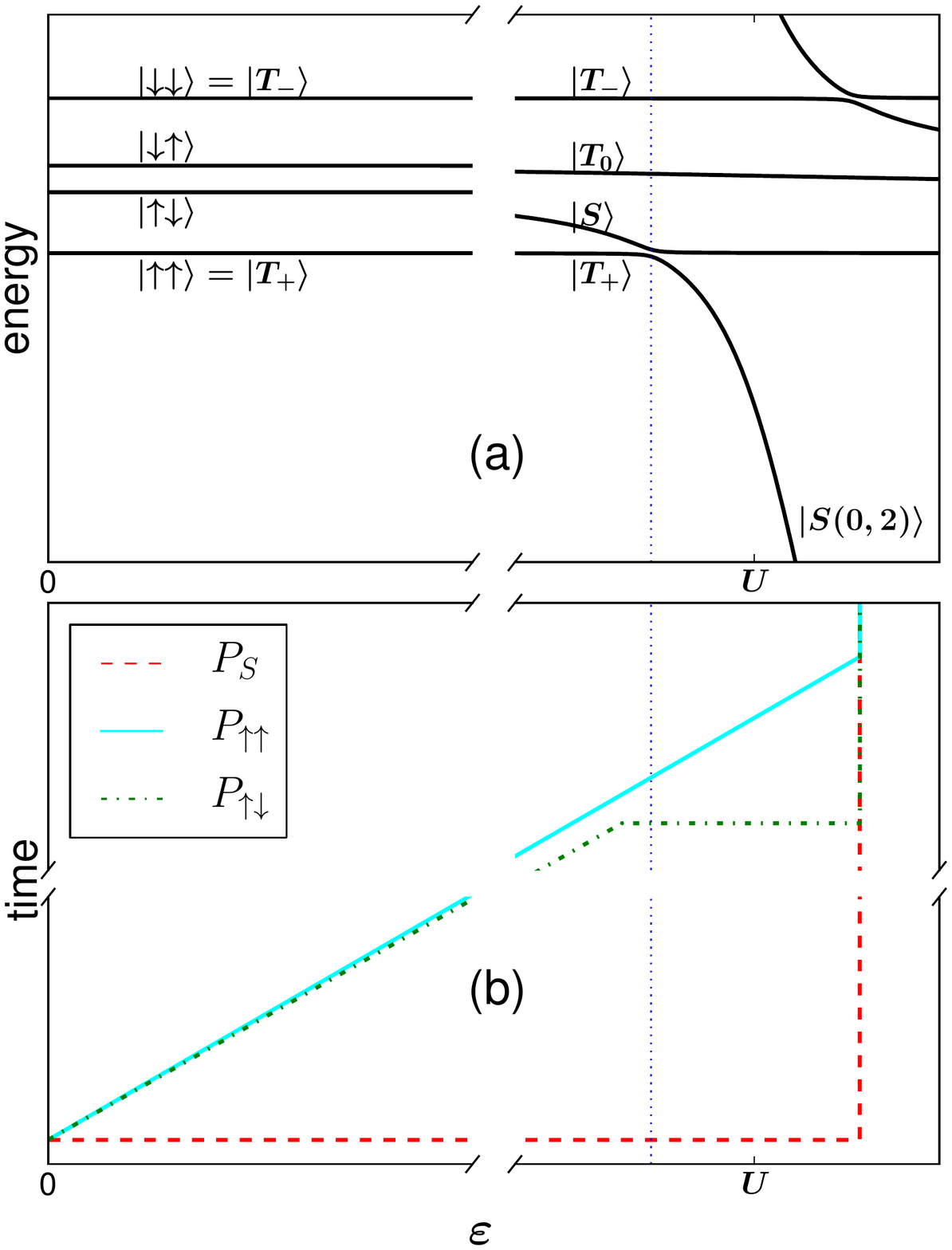}
\caption{(Color online) (a): Spectrum of a two-electron double quantum dot
with a magnetic field with $|h_z|\gg|\Delta h_z|\gg4t^2/U$, $h_z$, $\Delta h_z < 0$
for small values of $\varepsilon$ and $\varepsilon$ close to $U$.
The avoided crossing between $\ket{S}$ and $\ket{\uparrow\uparrow}$ at the blue dotted line
is induced by slightly different directions of the magnetic fields in the
left and the right dot.
(b): Three ways to tune $\varepsilon$ to achieve projection on
$\ket{\uparrow\downarrow}$, dash-dotted (green) line,
on $\ket{S}$, dashed (red) line,
and on $\ket{\uparrow\uparrow}$, solid (cyan) line.
Note that the $\varepsilon$-axis and the time-axis are broken}
\label{fig:spectrum}
\end{figure}
Let us restrict the discussion first to a magnetic field only in $z$-direction,
$\textbf{h}_1 = h_{1z}\textbf{e}_z$, $\textbf{h}_2 = h_{2z}\textbf{e}_z$.
The homogeneous part of the magnetic field, $h_z=\frac{h_{z1}+h_{z2}}{2}$, splits off the triplet states, because
$\ket{T_+} = \ket{\uparrow\uparrow}$
and
$\ket{T_-} = \ket{\downarrow\downarrow}$
obtain the Zeeman energy $\pm 2h_z$ whereas
$\ket{T_0} = \frac{\ket{\uparrow\downarrow}+\ket{\uparrow\uparrow}}{\sqrt{2}}$
is not affected.
A gradient in the magnetic field, described by $\Delta h_z = h_{1z}-h_{2z}$,
leads to a mixing of $\ket{S}$ and $\ket{T_0}$.
If $|\Delta h_z|$ and $|h_z|$ are large compared to $4t^2/U$,
which is the value of $J$ for $\varepsilon=0$,
then the energy eigenstates with charge configuration (1,1) are approximately
$\ket{T_+}$, $\ket{T_-}$, $\ket{\uparrow\downarrow}$, $\ket{\downarrow\uparrow}$.
At higher values of $\varepsilon$, $\ket{S}$ becomes an energy eigenstate,
see Fig.~\ref{fig:spectrum}.
Different directions in the magnetic field in the left and the right quantum dot can
additionally  induce an avoided crossing between $\ket{\uparrow\uparrow}$
and $\ket{S}$.

Projective measurements on the two-qubit space are done by charge measurements
after a spin-to-charge conversion,\cite{kane_nature,vandersypen}
which we describe in the following.
For $\varepsilon=0$ the four lowest energy levels belong to states with charge configuration (1,1),
which constitute the logical subspace of the system.
The starting point is an arbitrary state $\ket{\psi}$ in this logical subspace and $\varepsilon=0$.
By changing gate voltages, $\varepsilon$ can be increased to a value higher than $U$,
modifying the spectrum to a situation where the state $(0,2)$ is the ground
state of the system, see Fig.~\ref{fig:spectrum} (a).
Performing this varying of $\varepsilon$ backwards in time would map the (0,2) charge
state on one specific state in the logical subspace, $\ket{\psi_s}$.
The forward procedure thus converts the $\ket{\psi_s}$ amount of $\ket{\psi}$,
given by $\braket{\psi_s}{\psi}\ket{\psi_s}$, to the (0,2) charge state and the
remaining contributions of $\ket{\psi}$ are still in charge configuration (1,1).
Therefore, measuring a charge state (0,2), corresponds to a projection onto the state $\ket{\psi_s}$
in the logical subspace.
Now, we will consider three different cases shown in Fig.~\ref{fig:spectrum} (b)
and explain which is the state $\ket{\psi_s}$ for the different time dependences
of $\varepsilon$.
Assuming we are in the situation shown in Fig.~\ref{fig:spectrum} (a),
then a slow change of $\varepsilon$ from 0 to a value higher than $U$ would be an adiabatic
transition of the $\ket{T_+}$ state to the (0,2) state.
If we increase $\varepsilon$ slowly from 0 up to a value below the avoided crossing between $\ket{S}$ and
$\ket{T_+}$ and then proceed fast through this avoided crossing, the charge readout
effectively projects onto the state $\ket{\uparrow\downarrow}$.
An overall fast transition yiels a projection onto $\ket{S}$.
We will use this three different kinds of projection in our scheme for state tomography in Sec.~\ref{sec:quorum}.
In the state-tomography experiment for the $S-T_0$ qubits\cite{foletti2009,shulman} the projections onto
$\ket{\uparrow\downarrow}$ and onto $\ket{S}$ were used.
To allow projections onto other than the three states described so far, we assume quantum gates to be applied prior to the sweep.
In principle more complicated transitions are possible, they are described by a Landau-Zener transition.\cite{hugo_prl,hugo_prb}
State preparation is done in the opposite direction by converting the (0,2) state in a well-defined
(1,1) charge state.

It has been shown that the exchange interaction together with arbitrary single-qubit rotations
allows for universal quantum computing,\cite{LoDi1998} i.e., every unitary operation can be
constructed from these elementary gates.
To achieve full control on a single qubit, a tunable magnetic field in one direction is necessary
with individual tuning for each dot.
Additionally, rotations on the Bloch sphere around another axis than the $z$-axis
must be possible, which can be done using ESR.
This requires an oscillating magnetic field orthogonal to the axis of the time-independent
magnetic field.
The frequency of the oscillation is given by the Rabi frequency of the constant field.
As a sufficiently strong high-frequency magnetic field is hard to realize experimentally,
Brunner \textit{et al.}\cite{brunner} used an inhomogeneous magnetic field and
achieved an effective time-dependence in the Zeeman Hamiltonian by applying an ac voltage
shifting the electron back and forth.
Due to an offset in Zeeman energy between dots it was possible to perform ESR on each of the dots
separately.
Since exchange interaction was performed as well, in this double dot sample,
operations allowing for full control on the two-qubit space have been demonstrated
experimentally.

\section{General considerations about two-qubit tomography}
\label{sec:general}
We want to find a scheme for state tomography for the two-qubit space.
This means the scheme has to provide the full information of the density
matrix $\rho$, which is a Hermitian $4\times4$ matrix with
trace 1.
Therefore, we need at least 15 different measurements to determine
the 15 real parameters describing $\rho$.
In the space of $4\times4$ matrices we can introduce the scalar product
$\braket{A}{B}_M:= \operatorname{tr}(A^\dagger B)$ and the basis
$\{D_k{=}\frac{\sigma_{1i}\sigma_{2j}}{2} \text{ with } k{=}4i+j \text{ and }i,j\in\{0,x{\equiv}1,y{\equiv}2,z{\equiv}3\}\}$
where $\sigma_{1j}:=\sigma_j\otimes\mathbbm{1}$ and
$\sigma_{2j}:=\mathbbm{1}\otimes\sigma_j$ are Pauli matrices for
the first or second spin, and $\sigma_{n0}=\mathbbm{1}$.
Note that this basis is orthonormal with
respect to $\braket{\cdot}{\cdot}_M$ because
$\braket{\sigma_{1i}\sigma_{2j}}{\sigma_{1k}\sigma_{2l}}_M=4\delta_{ij}\delta_{kl}$.
We can expand $\rho$ in the $\{D_k\}$ basis,
\begin{equation}
\label{eqn:rho}
 \rho=\sum_{k=0}^{15}\rho_{k}D_k.
\end{equation}
We know already that all $\rho_{k}$ are real because $\rho$ is Hermitian
and that $\rho_{0}=1/2$ due to the trace condition.
The aim of the tomography is to determine every parameter $\rho_{k}$, $k=1,\ldots,15$.

Estimations for $\rho_1,\ldots,\rho_{15}$ have to be acquired
from experimental data.
In the physical setup we consider in this paper, measurements are done
by measuring the charge state after increasing the detuning $\varepsilon$.
This converts a state from the logical spin space (qubit space) with charge
configuration (1,1) to a (0,2) charge state.
In the previous section we found conditions under which this leads
to a projective measurement on either $\ket{\uparrow\downarrow}$, $\ket{S}$,
or $\ket{T_+}$.
We assume that universal quantum gates are feasible and therefore, any other
state could have been mapped by a unitary operation onto the state which is
converted to the (0,2) charge state.
Therefore, spin-to-charge conversion and universal quantum gates allow for
a projection onto an arbitrary state $\ket{\phi}$.
Each measurement within the tomography scheme will project onto a state
$\ket{\phi_j}$ and thus yields $\text{tr}(\ket{\phi_j}\bra{\phi_j}\rho)$.
The projection operator $P_j=\ket{\phi_j}\bra{\phi_j}$ is also a
Hermitian $4\times4$ matrix with trace 1, which means
$\braket{P_j}{D_0}_M=\frac{1}{2}\braket{P_j}{\mathbbm{1}}_M=1/2$.
Nevertheless, $P_j$ can be expressed in terms of the six real parameters\footnote{Note that
a state in a four-dimensional Hilbert space is given by four complex, i.e.,
eight real parameters, but two of them are fixed here due to normalization
and the irrelevance of a global phase.}
for the state $\ket{\phi_j}$.
For $P_j$ given in the $\frac{\sigma_{1i}\sigma_{2j}}{2}$ basis, see Appendix
\ref{app:projector}.
We are seeking a set $\{\ket{\phi_j}\}$ that provides full information
of $\rho$.
If 15 projectors fulfill the condition that
$\{P_j-\mathbbm{1}/4, j{=}1,\ldots,15\}$ is a basis in the space of traceless
$4\times4$ matrices, we call these projectors a minimal set or quorum.

We introduce the invertible $15\times15$ matrix $\mathcal{P}$ with
\begin{equation}
 \mathcal{P}_{jk} = \braket{P_j}{D_k}_M \text{ with } j,k=1,\ldots,15,
\end{equation}
i.e., considering only the traceless contributions of the projectors $P_j$.
Since $\operatorname{tr}(P_j\rho)=\frac{1}{4}+\sum_{k=1}^{15}\mathcal{P}_{jk}\rho_k$,
the parameters $\rho_k$ are given by $\rho_k = \sum_{j=1}^{15} \mathcal{P}^{-1}_{kj}(\operatorname{tr}(P_j\rho)-\frac{1}{4})$.
We obtain an experimental estimation $m_j$ for $\operatorname{tr}(P_j\rho)$ by
dividing the number of the charge state (0,2) outcomes by the number of
experimental runs for the $j$th projection experiment.
This statistical data does not yield the exact values of $\rho_k$ but
estimations for it, which we call $\tilde\rho_k$ and which are by linear reconstruction given by  
\begin{equation}
\label{eqn:rec}
 \left(\begin{array}{c}\tilde\rho_{1}\\\vdots\\\tilde\rho_{15}\end{array}\right) =
\mathcal{P}^{-1}\left(\begin{array}{c}m_1-\frac{1}{4}\\\vdots\\m_{15}-\frac{1}{4}\end{array}\right).
\end{equation}
Although the maximum likelihood method would be preferred over the linear
reconstruction for practical purposes we want to consider Eq.~(\ref{eqn:rec})
in order to obtain information about the influence of $\mathcal{P}$ on
the statistical error expressed by the covariance matrix
$
 C_{kl} =E((\tilde\rho_{l}-\rho_{l})(\tilde\rho_{k}-\rho_{k}))
$
where $E(\cdot)$ is the expectation value of the probability distribution, which is
given by the density matrix $\rho$ and the set of measurements $P_1,\ldots,P_{15}$.
Note that $E(\cdot)$ is in general not expressed by the quantum mechanical expectation
value $\operatorname{tr}(O\rho)$ of an operator $O$,
but this is the case for those operators which are directly measured, i.e.,
for every $j=1,\ldots,15$ we have a binomial distribution with the number
of trials, $N_j$, and the probability of success $\operatorname{tr}(P_j\rho)$.
The probability distribution for the indirectly obtained stochastic variables
$\tilde\rho_k$ has to be gained using the matrix $\mathcal{P}$.
Using Eq.~(\ref{eqn:rec}), we find for the covariance matrix
\begin{equation}
\begin{split}
 C_{kl} &= E\left(\sum_{j=1}^{15}\mathcal{P}^{-1}_{kj}(m_j-\operatorname{tr}(P_j\rho))\sum_{i=1}^{15}(m_i-\operatorname{tr}(P_i\rho))\mathcal{P}^{-1}_{li}\right)\\
               &= \sum_{i,j=1}^{15}\mathcal{P}^{-1}_{kj}\underbrace{E((m_j-\operatorname{tr}(P_j\rho))(m_i-\operatorname{tr}(P_i\rho)))}_{=B_{ji}}\mathcal{P}^{-1}_{li},\\
\end{split}
\end{equation}
which is equivalent to $C=\mathcal{P}^{-1}B(\mathcal{P}^{-1})^T$.
Our aim is to find a set of measurement for which every value $|C_{kl}|$ is small.
The covariance matrix of the measurement results, $B$, is given by 
$B_{ji}=\delta_{ji}\operatorname{tr}(P_j\rho)(1-\operatorname{tr}(P_j\rho))/N_j\leq \delta_{ji}/(4N_j)$
as a property of the binomial distribution describing $m_j$,
and the upper limit is reached if $\operatorname{tr}(P_j\rho)=1/2$.
Using the adjugate matrix,
$(\operatorname{adj}\mathcal{P})_{kl}=(-1)^{k+l}\det(\mathcal{P}_{/l/k})$
where $\mathcal{P}_{/l/k}$ is the matrix one gains by deleting the $l$th
row and the $k$th column of $\mathcal{P}$,
to express the inverse of $\mathcal{P}$ we obtain
\begin{equation}
 |C_{kl}| = \left|\sum_{j=1}^{15} \mathcal{P}^{-1}_{kj}\mathcal{P}^{-1}_{lj} \frac{\operatorname{tr}(P_j\rho)(1-\operatorname{tr}(P_j\rho))}{N_j}\right|
          \leq \sum_{j=1}^{15} \frac{|(\operatorname{adj}\mathcal{P})_{kj}(\operatorname{adj}\mathcal{P})_{lj}|}{4N_j\det(\mathcal{P})^2}.
\end{equation}
The \textit{length} in matrix space of $P_j-\mathbbm{1}/4$, i.e. a row of $\mathcal{P}$, is
$\sqrt{\braket{ P_j-\frac{\mathbbm{1}}{4}}{P_j-\frac{\mathbbm{1}}{4}}_M} = \sqrt{\frac{3}{4}}$.
A determinant is invariant under the Gram-Schmidt orthogonalization,
yielding upper limits for $|\det(\mathcal{P})|$ as well as for
the matrix elements of the adjugate matrix of $\mathcal{P}$, which are
determinants of $14\times14$ matrices gained from $\mathcal{P}$ by
deleting one row and one column.
For $\operatorname{adj}\mathcal{P}$ we find
$|(\operatorname{adj} \mathcal{P})_{kj}|\leq (\frac{3}{4})^{\frac{14}{2}}$.
Assuming $N_1=N_2=\cdots=N_{15}=N$, this results in the following inequality
\begin{equation}
 |C_{kl}| \leq \frac{15(\frac{3}{4})^{14}}{4N\det(\mathcal{P})^2}\approx\frac{0.06682}{N\det(\mathcal{P})^2}.
\end{equation}
We now concentrate on maximizing the value $\det(\mathcal{P})^2$.
The upper bound for $|\det(\mathcal{P})|$ is given by
$(\frac{3}{4})^{\frac{15}{2}}\approx0.1156$, which would be reached if the
rows of $\mathcal{P}$ were orthogonal, which means the 15 projectors
$P_j$ have to fulfill
\begin{equation}
\label{eqn:orth_in_matrix_space}
 \braket{P_j-\mathbbm{1}/4}{P_k-\mathbbm{1}/4}_M = 0~\Leftrightarrow~\braket{P_j}{P_k}_M=|\braket{\phi_j}{\phi_k}|^2=1/4\text{ for }j\neq k.
\end{equation}
In Appendix \ref{app:proof} we prove that this is not possible.

The quorum from Ref.~\onlinecite{james} for the two-qubit space, after
translation from photon polarization states to spin states, reads
$\{\ket{\uparrow\uparrow}$,
$\ket{\uparrow\downarrow}$,
$\ket{\downarrow\uparrow}$,
$\ket{\downarrow_y\uparrow}$,
$\ket{\downarrow_y\downarrow}$,
$\ket{\uparrow_x\downarrow}$,
$\ket{\uparrow_x\uparrow}$,
$\ket{\uparrow_x\downarrow_y}$,
$\ket{\uparrow_x\uparrow_x}$,
$\ket{\downarrow_y\uparrow_x}$,
$\ket{\uparrow\uparrow_x}$,
$\ket{\downarrow\uparrow_x}$,
$\ket{\downarrow\uparrow_y}$,
$\ket{\uparrow\uparrow_y}$,
$\ket{\downarrow_y\uparrow_y}\}$
where $\ket{\uparrow_y}=\frac{\ket{\uparrow}+i\ket{\downarrow}}{\sqrt{2}}$,
$\ket{\downarrow_y}=\frac{\ket{\uparrow}-i\ket{\downarrow}}{\sqrt{2}}$, and
$\ket{\uparrow_x}=\frac{\ket{\uparrow}+\ket{\downarrow}}{\sqrt{2}}$.
We have removed the state $\ket{\downarrow\downarrow}$ because 15 states are
sufficient here.
These basis states are all separable and lead
to $|\det(\mathcal{P})|=\frac{1}{512}\approx2\cdot10^{-3}$.

A better result can be obtained by the well-known concept of mutually
unbiased bases.\cite{ivanovic,durt}
Two bases in Hilbert space are called mutually unbiased if a state of the first
basis has the same overlap with all states of the second basis.
The idea of mutually unbiased bases originally referred to a situation where $N+1$ observables
are measured,\cite{ivanovic} and each of this measurements yields the probability for being in one of the
corresponding $N$ eigenstates, which are here chosen to be basis states of one of the mutually
unbiased bases.
In that case it was shown that the mutually unbiased bases represent an optimal set to determine
$\rho$ in the sense of minimizing the statistical error.\cite{wootters_fields}
For the projective measurement by the spin-to-charge conversion, which we consider in this paper,
the situation is slightly different since every projection is done by another measurement.
In particular, it is not necessary to measure all 20 probabilities; it suffices to perform the
projective measurement for three states out of the four states in each of the five bases,
providing a basis in the 15 dimensional matrix space, and consequently a minimal set of projectors.
It has been shown  that for a $N$-level system a set of
$N+1$ mutually unbiased bases exist if $N$ is the integer power of a prime
number.\cite{wootters_fields}
Hence, for our four-dimensional case, five bases
$\{\ket{\phi_{0,1}},\ket{\phi_{0,2}},\ket{\phi_{0,3}},\ket{\phi_{0,4}}\},\ldots,\{\ket{\phi_{4,1}},\ldots,\ket{\phi_{4,4}}\}$
fulfilling
\begin{equation}
 |\braket{\phi_{jk}}{\phi_{lm}}|^2 = \delta_{jl}\delta_{km} + \frac{1-\delta_{jl}}{4}.
\end{equation}
can be found.
Note that states from different bases with numbers $j\neq l$  fulfill
$|\braket{\phi_{jk}}{\phi_{lm}}|^2=\frac{1}{4}$, which is the condition
(\ref{eqn:orth_in_matrix_space}). This means that the matrix space spanned
by the traceless parts of the corresponding projectors contains five
three-dimensional orthogonal subspaces.
On the other hand, the states in the same basis are orthogonal in
Hilbert space, $\braket{\phi_{jk}}{\phi_{jm}}=\delta_{mk}$.
We show in Appendix \ref{app:mub} that this always leads to
$|\det(\mathcal{P})|=\frac{1}{32}=0.03125$.

\section{Explicit construction of spin-qubit quorum}
\label{sec:quorum}
We use a set of mutually unbiased bases given in Ref.~\onlinecite{klappenecker_roetteler}
and present an explicit construction of the following 15 states out of these bases, forming a
quorum,
\begin{equation}
\label{eqn:quorum}
 \begin{split}
& \ket{\psi_1} = \ket{\uparrow\uparrow},  \hspace{1cm} P_1=\frac{\mathbbm{1}+\sigma_{1z}+\sigma_{2z}+\sigma_{1z}\sigma_{2z}}{4},\\
& \ket{\psi_2} = \ket{\uparrow\downarrow},  \hspace{1cm} P_2=\frac{\mathbbm{1}+\sigma_{1z}-\sigma_{2z}-\sigma_{1z}\sigma_{2z}}{4},\\
& \ket{\psi_3} = \operatorname{SWAP} \ket{\uparrow\downarrow} = \ket{\downarrow\uparrow},  \hspace{1cm} P_3=\frac{\mathbbm{1}-\sigma_{1z}+\sigma_{2z}-\sigma_{1z}\sigma_{2z}}{4},\\
& \ket{\psi_4} = e^{i\frac{\pi}{2}\sigma_{2 z}}\sqrt{\operatorname{SWAP}}e^{i\frac{\pi}{4}\sigma_{1 x}}\ket{S}=\ket{\uparrow_x\uparrow_x},
  \hspace{1cm} P_4=\frac{\mathbbm{1}+\sigma_{1x}+\sigma_{2x}+\sigma_{1x}\sigma_{2x}}{4},\\
& \ket{\psi_5} = e^{i\frac{\pi}{2}\sigma_{1 z}}\ket{\psi_4}=\ket{\downarrow_x\uparrow_x},
  \hspace{1cm} P_5=\frac{\mathbbm{1}-\sigma_{1x}+\sigma_{2x}-\sigma_{1x}\sigma_{2x}}{4},\\
& \ket{\psi_6} = e^{i\frac{\pi}{2}\sigma_{2 z}}\ket{\psi_4}=\ket{\uparrow_x\downarrow_x},  \hspace{1cm} P_6=\frac{\mathbbm{1}+\sigma_{1x}-\sigma_{2x}-\sigma_{1x}\sigma_{2x}}{4},\\
& \ket{\psi_7} = e^{-i\frac{\pi}{4}(\sigma_{1 z} + \sigma_{ 2 z})}\ket{\psi_4}=\ket{\uparrow_y\uparrow_y},
  \hspace{1cm} P_7=\frac{\mathbbm{1}+\sigma_{1y}+\sigma_{2y}+\sigma_{1y}\sigma_{2y}}{4},\\
& \ket{\psi_8} = e^{i\frac{\pi}{2}\sigma_{1 z}}\ket{\psi_7}=\ket{\downarrow_y\uparrow_y},
  \hspace{1cm} P_8=\frac{\mathbbm{1}-\sigma_{1y}+\sigma_{2y}-\sigma_{1y}\sigma_{2y}}{4},\\
& \ket{\psi_9} = e^{i\frac{\pi}{2}\sigma_{2 z}}\ket{\psi_7}=\ket{\uparrow_y\downarrow_y},
  \hspace{1cm} P_9=\frac{\mathbbm{1}+\sigma_{1y}-\sigma_{2y}-\sigma_{1y}\sigma_{2y}}{4},\\
& \ket{\psi_{10}} = e^{i\frac{\pi}{4}\sigma_{1 x}}e^{i\frac{\pi}{8}(\sigma_{1 z} - \sigma_{2 z})}\ket{S},
  \hspace{1cm} P_{10}=\frac{\mathbbm{1}-\sigma_{1z}\sigma_{2x}-\sigma_{1x}\sigma_{2y}-\sigma_{1y}\sigma_{2z}}{4},\\
& \ket{\psi_{11}} = e^{i\frac{\pi}{2}\sigma_{1 z}} \ket{\psi_{10}},
  \hspace{1cm} P_{11}=\frac{\mathbbm{1}-\sigma_{1z}\sigma_{2x}+\sigma_{1x}\sigma_{2y}+\sigma_{1y}\sigma_{2z}}{4},\\
& \ket{\psi_{12}} = e^{i\frac{\pi}{2}\sigma_{2 z}}\ket{\psi_{10}},
  \hspace{1cm} P_{12}=\frac{\mathbbm{1}+\sigma_{1z}\sigma_{2x}+\sigma_{1x}\sigma_{2y}-\sigma_{1y}\sigma_{2z}}{4},\\
& \ket{\psi_{13}} = e^{-i\frac{\pi}{4}(\sigma_{1 z} + \sigma_{ 2 z})}\ket{\psi_{10}},
  \hspace{1cm} P_{13}=\frac{\mathbbm{1}+\sigma_{1y}\sigma_{2x}-\sigma_{1z}\sigma_{2y}+\sigma_{1x}\sigma_{2z}}{4},\\
& \ket{\psi_{14}} = e^{i\frac{\pi}{2}\sigma_{1 z}}\ket{\psi_{13}},
  \hspace{1cm} P_{14}=\frac{\mathbbm{1}-\sigma_{1y}\sigma_{2x}-\sigma_{1z}\sigma_{2y}-\sigma_{1x}\sigma_{2z}}{4},\\
& \ket{\psi_{15}} = e^{i\frac{\pi}{2}\sigma_{2 z}}\ket{\psi_{13}},
  \hspace{1cm} P_{15}=\frac{\mathbbm{1}-\sigma_{1y}\sigma_{2x}+\sigma_{1z}\sigma_{2y}+\sigma_{1x}\sigma_{2z}}{4},
 \end{split}
\end{equation}
where the
states $\ket{\psi_{3i+1}},\ket{\psi_{3i+2}},\ket{\psi_{3i+3}}$ are belonging to the same
basis, $i=0,1,2,3,4$.
The first three bases ($i=0,1,2$) are given by the product of spin states in
each dot where the quantization axis is chosen to be $z$, $x$, and
$y$, thus those states are separable.
The other two bases ($i=3,4$) consist of maximally entangled states.

\begin{figure}
 \begin{minipage}{0.4\textwidth}
  \includegraphics[width=4cm, angle=270]{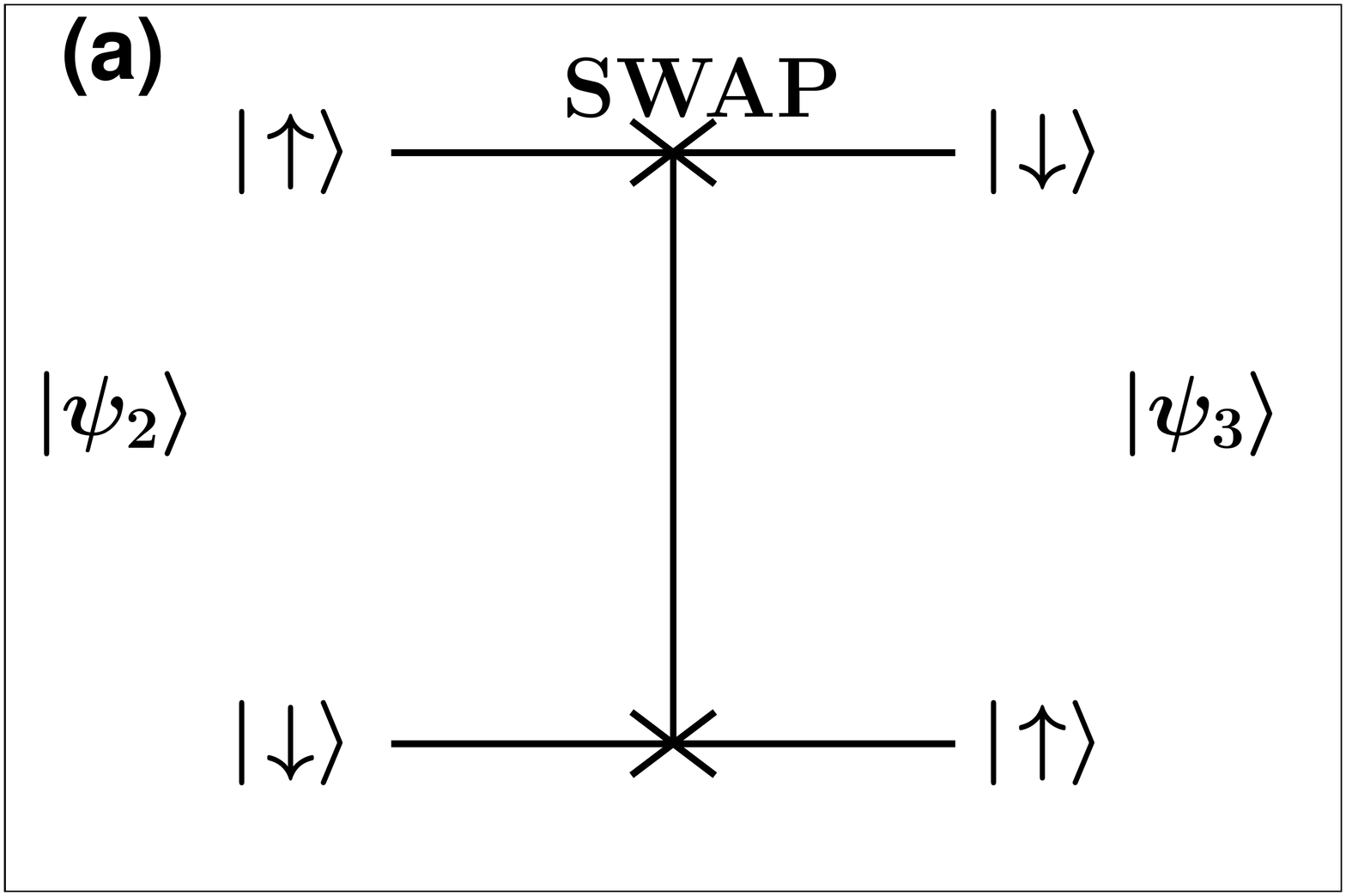}
 \end{minipage}
 \begin{minipage}{0.4\textwidth}
  \includegraphics[width=4cm, angle=270]{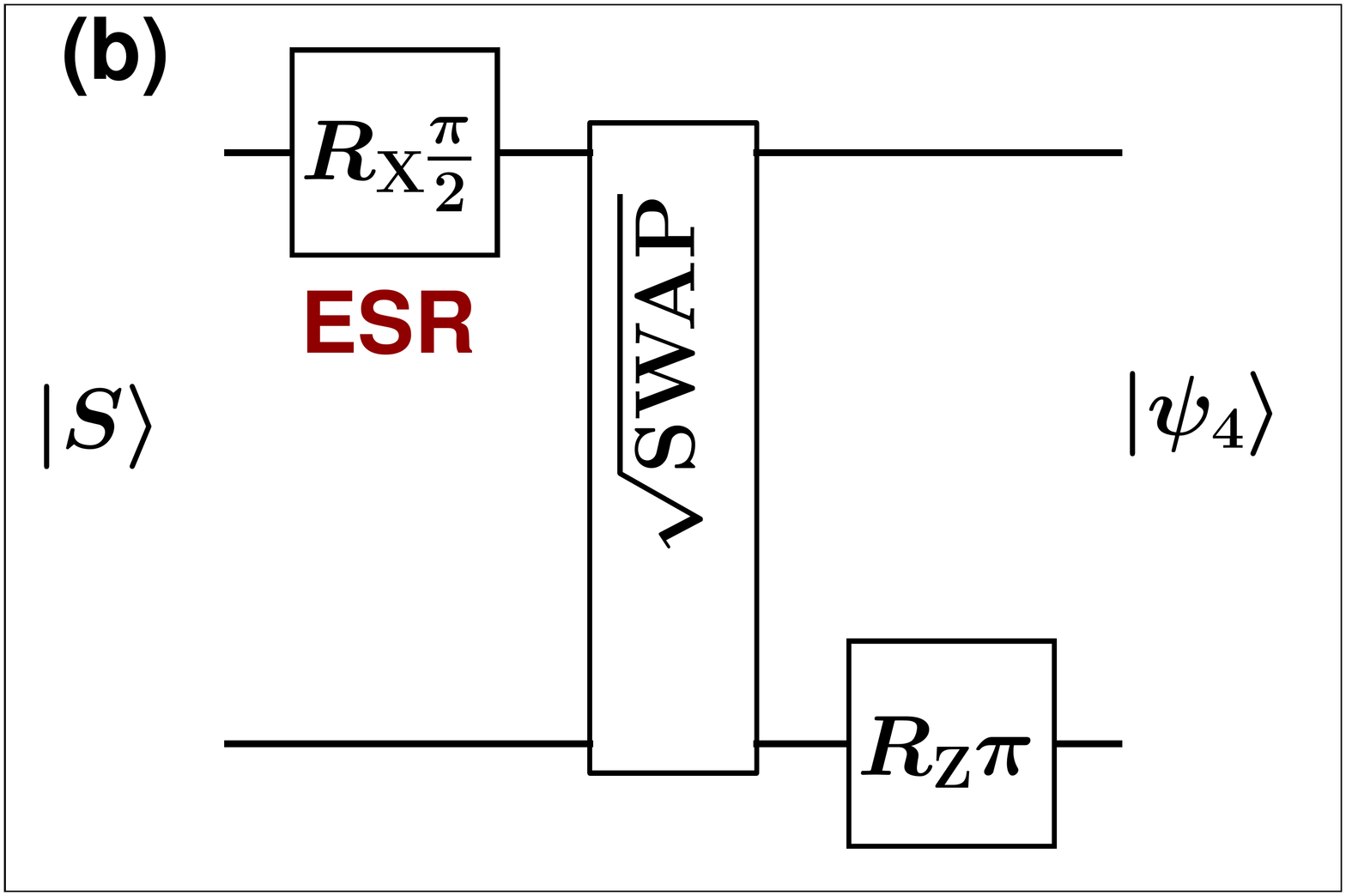}
 \end{minipage}
 \begin{minipage}{0.4\textwidth}
  \includegraphics[width=4cm, angle=270]{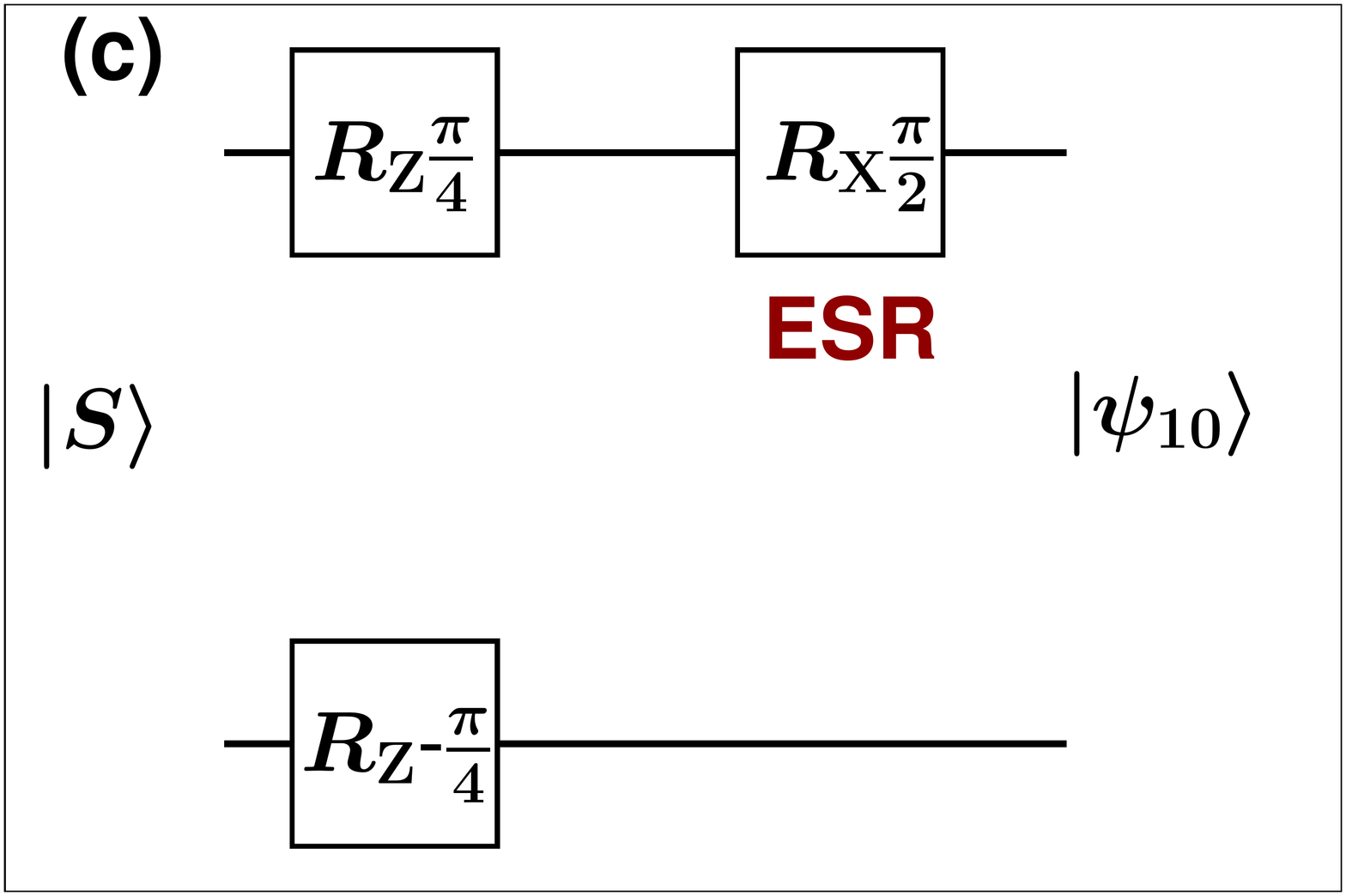}
 \end{minipage}
 \begin{minipage}{0.4\textwidth}
  \includegraphics[width=4cm, angle=270]{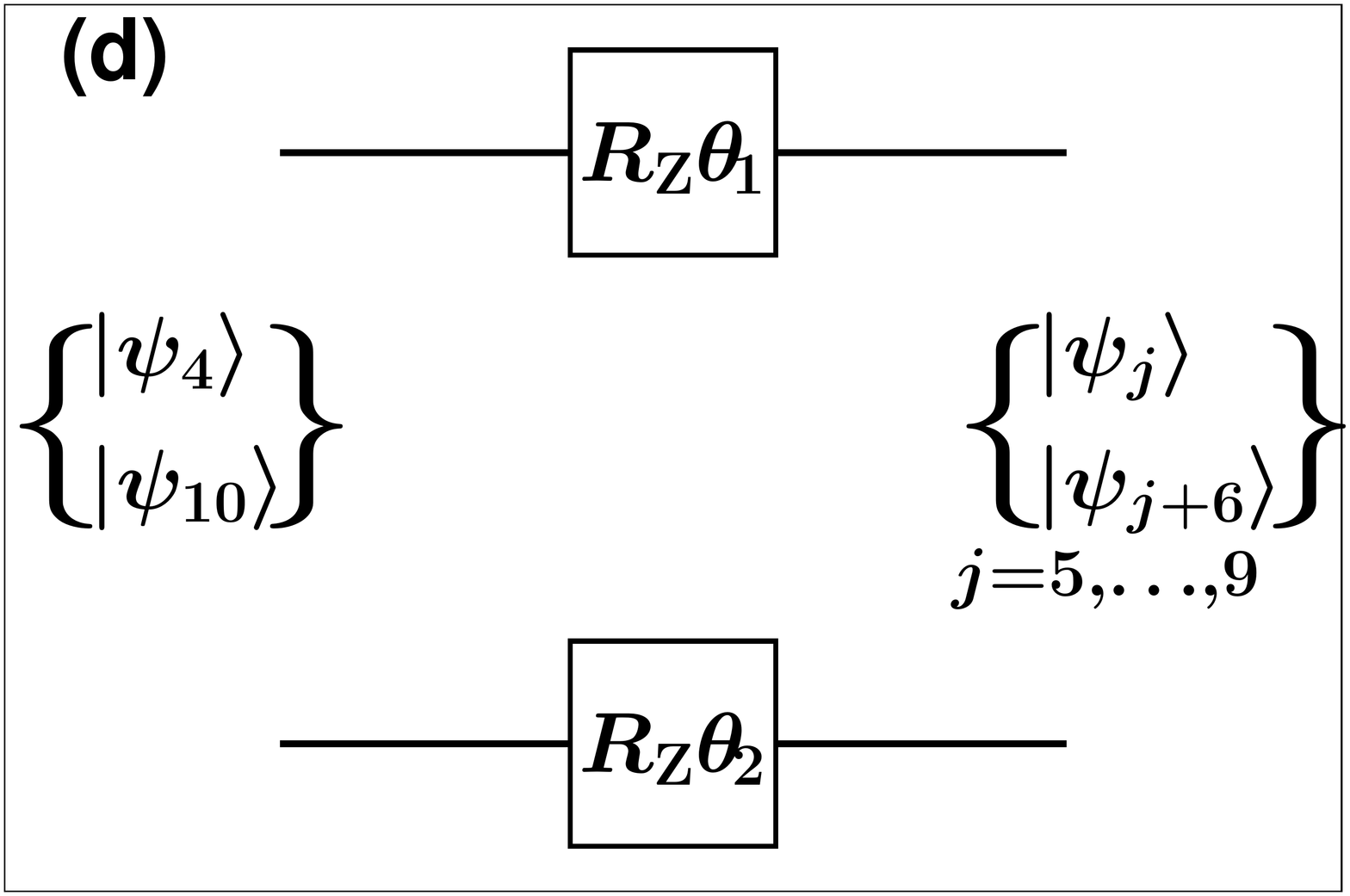}
 \end{minipage}
\caption{The quantum circuits used to realize the quorum (\ref{eqn:quorum}).
Note that ESR is used in the preparation of the states $\ket{\psi_4}$
and $\ket{\psi_{10}}$, the states $\ket{\psi_5},\ldots,\ket{\psi_9}$
and $\ket{\psi_{11}},\ldots,\ket{\psi_{15}}$ can then be gained by rotations
due to a static magnetic field from $\ket{\psi_4}$ or $\ket{\psi_{10}}$
respectively.
The gate $R_X\frac{\pi}{2}$ on the upper line in panel (b)
denotes a rotation of the first qubit around the $x$-axis by the angle
$\pi/2$, i.e., the operation $e^{i\frac{\pi}{4}\sigma_{1x}}$.
The other rotations $R_X$ and $R_Z$ are defined accordingly.
The angles $\theta_1$ and $\theta_2$ in panel (d) depend on the index $j$
of the final state but they are the same for the maps $\ket{\psi_4}\mapsto\ket{\psi_j}$
and $\ket{\psi_{10}}\mapsto\ket{\psi_{j+6}}$.
The actual values of $\theta_1$ and $\theta_2$ can be found in Eq.~(\ref{eqn:quorum}), e.g.,
for $j=9$ they are given by $\theta_1=-\pi/2$ and $\theta_2=\pi/2$.}
\label{fig:gates}
\end{figure}

Within the scheme we are using projective measurements on the states
 $\ket{\uparrow\uparrow}$, $\ket{\uparrow\downarrow}$, and $\ket{S}$
as described in Sec.~\ref{sec:phys_sys}.
In order to access the states in the set above not directly given by one
of these states, we apply various quantum gates directly available for spin qubits,
see Fig.~\ref{fig:gates}.
The ${\rm SWAP}$ gate is defined by ${\rm SWAP}\ket{s_1s_2}=\ket{s_2s_1}$,
i.e., it interchanges the information of the first and the second qubit.
It can be realized by applying the exchange interaction (\ref{eqn:exch}) for
a time $t_e$ given by $\varphi=\int_0^{t_e}J(\tau)d\tau = \pi$,
where we neglected the magnetic field (gradient) which is possible if
$J$ is large compared to the gradient $\Delta h_z$.
Consequently, the $\sqrt{\operatorname{SWAP}}$ gate can be achieved by applying
the exchange interaction for yielding $\varphi=\pi/2$.
Since $\sigma_{1z}+\sigma_{2z}$ commutes with the exchange coupling, the
influence of $h_z$ can be treated formally as if it was applied before
or after the exchange interaction.
Furthermore, we need an ESR pulse on the first spin to access the states
$\ket{\psi_4},\ldots,\ket{\psi_{15}}$.
We want to emphasize that for each of these twelve states we need only one
ESR $\pi/2$ rotation, $e^{i\frac{\pi}{4}\sigma_{1x}}$ of only one of the spins.
As we assume the fidelity of ESR operations to be lower than that of
exchange gates or rotations around the $z$-axis, this is a crucial point
for an experimental realization of a tomography scheme in a double quantum dot.

Note that ESR cannot be completely avoided because exchange interaction and
rotations due to a static magnetic field together with the projections on
$\ket{\uparrow\downarrow}$, $\ket{S}$, $\ket{\uparrow\uparrow}$ can only
access the five-dimensional subspace in matrix space spanned by
$\{\sigma_{1z},\sigma_{2z},\sigma_{1z}\sigma_{2z},\sigma_{1x}\sigma_{2x}+\sigma_{1y}\sigma_{2y},\sigma_{1x}\sigma_{2y}-\sigma_{1y}\sigma_{2x}\}$.
This can be seen by regarding the time evolution under exchange interaction,
which leaves $P_{\uparrow\uparrow}=P_1$ and $P_S$ invariant and evolves
$P_{\uparrow\downarrow}=P_2$ to
\begin{equation}
 e^{i\varphi P_S}P_{\uparrow\downarrow}e^{-i\varphi P_S} =
\frac{\mathbbm{1}+\sigma_{1z}\sigma_{2z}+\cos\varphi(\sigma_{1z}-\sigma_{2z})+\sin\varphi(\sigma_{1x}\sigma_{2y}-\sigma_{1y}\sigma_{2x})}{4},
\end{equation}
and the time evolution under a gradient Zeeman field in $z$-direction,
leaving $P_{\uparrow\downarrow}$ and $P_{\uparrow\uparrow}$ invariant
and evolves $P_S$ to
\begin{equation}
 e^{i\frac{\vartheta}{4}(\sigma_{1z}-\sigma_{2z})}P_Se^{-i\frac{\vartheta}{4}(\sigma_{1z}-\sigma_{2z})} =
\frac{\mathbbm{1}+\sigma_{1z}\sigma_{2z}+\cos\vartheta(\sigma_{1x}\sigma_{2x}\!+\!\sigma_{1y}\sigma_{2y})+\sin\vartheta(\sigma_{1x}\sigma_{2y}\!-\!\sigma_{1y}\sigma_{2x})}{4}.
\end{equation}

The single-qubit gates containing $\sigma_{1z}$ or $\sigma_{2z}$ can in
principle be achieved just by waiting for a certain time with negligible $J$
and an applied magnetic field.
Since the magnetic field cannot easily be switched off completely this requires
precise control of the time which is spent between state preparation and
the projective measurement. Considering for example the states $\ket{\psi_4}$
and $\ket{\psi_7}$, we see that the state with spins polarized in $x$-direction,
$\ket{\psi_4}$, and the one with both spins in $y$-direction, $\ket{\psi_7}$,
are connected just by a rotation around the $z$-axis for both spins, which is
realized by a (large) constant magnetic field.
But clearly it is necessary to distinguish those states $\ket{\psi_4}$ and
$\ket{\psi_7}$ experimentally to get the full information of the density matrix
$\rho$.
Therefore, state tomography requires more control than would be necessary to demonstrate
ESR rotations.
Note that the adjoint quantum gates included in (\ref{eqn:quorum}) are applied on
the initial state in the tomography experiment.
Afterward a lack of time control might not harm the experiment as a
$e^{i\phi(\sigma_{1z}+\sigma_{2z})}$ gate results in no more than a global phase on
the states $\ket{\uparrow\downarrow}$, $\ket{\uparrow\uparrow}$, and $\ket{S}$.
Only the projection on these states by the spin-to-charge conversion has to be precise.

\section{Estimating the error}
\label{sec:error}
Measuring a quantum system always produces statistical results, thus
the exact quantum state cannot be determined by an experiment which
is performed a finite number of times.
But increasing this number of repetitions cannot lead to an arbitrary high
accuracy of the estimation for the quantum state either since the measurement
scheme unavoidably contains systematic errors.
Therefore, it is necessary to know how precise the desired measurement scheme
is actually performed.
We assume the error in the measurement setup to arise mainly from the
applied quantum gates, and among those the ESR gate, $e^{i\frac{\pi}{4}\sigma_{1x}}$,
to have the largest influence an the total error.
In the experiments described in Ref.~\onlinecite{brunner}
the fidelity of the exchange rotation could not be extracted because
two ESR rotations were also included in the explored two-qubit
gate affecting strongly the fidelity of the combined gate.
Errors in the quantum gates may occur due to fluctuations in the magnetic field,
for instance, caused by nuclear spins,\cite{khaetskii,petta}
which is most damaging for the ESR gate. Charge fluctuations mainly influence
the exchange interaction and thus the $\operatorname{SWAP}$ and
$\sqrt{\operatorname{SWAP}}$ gate. Finally, differences in the times for which interactions
are actually switched on, affect all quantum gates but in particular the spin rotations
around the $z$-axis according to a static magnetic field.

Knowledge about the experimentally realized precision of the projective measurements
suggested in Sec.~\ref{sec:quorum}, can practically only gained by prior experiments.
One option to achieve this is a self-consistent tomography\cite{merkel,medford}
where the perfect states from Sec.~\ref{sec:quorum}, $P_j$ ($j=1,\ldots,15$), are replaced by
imperfect and unknown ones, $P_i'$. This implies that before using the set of measurements to
determine an unknown quantum state, the set $\{P_j'\}$ itself has to be determined first.
For this purpose, one could, e.g., perform measurements to obtain statistical knowledge on the quantities
\begin{equation}
 M_{ij} = \operatorname{tr}(P_i'P_j'),\hspace{2cm}i,j=1,\ldots 15.
\end{equation}
Note that preparing the system in the state $P_i'$ is the inverted procedure of
the projective measurement on this state, staring from a (0,2) charge state.
The states $P_j'$ can be determined approximately afterward
via a maximum likelihood algorithm\cite{merkel,medford}
up to a basis in Hilbert space which has to be fixed as $M_{ij}$ is
invariant under a unitary transformation.
Note that here, less information is needed than for quantum process tomography.\cite{merkel}

Since the 15 states in Sec.~\ref{sec:quorum} are generated by a smaller number of different
quantum gates it might be an alternative to describe those gates with a set of parameters
including mean fluctuations and determine these parameters experimentally.
In general, the imperfect projection operators $P_j'$ would be assumed to depend on a set of
$n$ parameters $\alpha_1,\ldots,\alpha_n$,
for a parametrization of fluctuating magnetic and electric fields see Ref.~\onlinecite{medford},
for a parameter fit of a ESR experiment see Ref.~\onlinecite{koppens}.
For experiments which are repeated often, we can consider the mixed states $\overline {P_j'}$
calculated from the statistical distribution of the parameters $\alpha_1,\ldots,\alpha_n$,
\begin{equation}
 \overline {P'_j}= \int d\alpha_1\ldots d\alpha_n\,p_1(\alpha_1)\ldots p_n(\alpha_n) P_j'(\alpha_1,\ldots,\alpha_n).
\end{equation}
For Gaussian distributions it is sufficient to determine the mean values of $\alpha_i$ and $\alpha_i^2$,
which means that $n^2$ numbers have to be extracted from experiments.
Note that errors in those values will directly lead to errors in the tomography of
an unknown state, but it is not necessary that the distributions are sharp.
Nevertheless, the deviations of the mixed states $P_j'$ from the pure states $P_j$
worsen the result of the tomography measurement as illustrated by the following example.
Assume the matrices $P_4'-\mathbbm{1}/4$ and $P_6'-\mathbbm{1}/4$ to have the same
directions in matrix space as $P_4-\mathbbm{1}/4$ and $P_6-\mathbbm{1}/4$, i.e.,
they are given by
\begin{equation}
 P_j' = \frac{1-f_j^2}{3}\mathbbm{1}+\frac{4f_j^2-1}{3}P_j,
\end{equation}
where $f_j=\operatorname{tr}\left(\sqrt{\sqrt{P_j}P_j'\sqrt{P_j}}\right)$ is
the fidelity of the quantum state $P_j'$ with respect to $P_j$,
and where we assume $f_4=f_6=f$ for the reminder of the discussion.
For a completely mixed state $P_j'=\mathbbm{1}/4$,
we find $f=1/2$.
Whereas $\rho_{4}$ from Eq.~(\ref{eqn:rho}) in the ideal case is given by
\begin{equation}
 \rho_{4} = \frac{2(\operatorname{tr}(P_4\rho)+\operatorname{tr}(P_6\rho))-1}{2},
\end{equation}
for the mixed states it is represented by
\begin{equation}
 \rho_{4} = \frac{3[2(\operatorname{tr}(P'_4\rho)+\operatorname{tr}(P'_6\rho))-1]}{2(4f^2-1)}.
\end{equation}
For the statistical error of $\rho_{4}$ desired to be smaller than $\delta$, the
statistical error of the experimental values for
$\operatorname{tr}(P'_4\rho)$ and $\operatorname{tr}(P'_6\rho)$, which we assume to be equal,
has to be below $\delta'=\frac{(4f^2-1)\delta}{3\sqrt{2}}$.
A Chernoff bound\cite{chernoff,hoeffding} can be applied to estimate the number
of experimental runs needed to realize a desired precision.\cite{bendersky_pastawski_paz,bendersky_paz}
The possible outcomes of one experimental run are 0 for no projection and 1 if
a (0,2) charge state is measured, the expectation value is $\operatorname{tr}(P'_j\rho)$.
Repeating this experiment $N_{\rm run}$ times thus leads to a binomial distribution.
The probability for the experimental value, i.e., the number of runs with result 1 divided
by $N_{\rm run}$, to be out of an interval
$[\operatorname{tr}(P'_j\rho)-\delta',\operatorname{tr}(P'_j\rho)+\delta']$,
$P_{\rm out}$, is limited by
\begin{equation}
 P_{\rm out} \le 2\exp{\left(\frac{-2\delta'^2}{N_{\rm run}}\right)}.
\end{equation}
Therefore, the number of experimental runs which provide a $P_{\rm out}$ below
a desired limit $P_l$ can be chosen by
\begin{equation}
 N_{\rm run} > \frac{\ln(2/P_l)}{2\delta'^2}=\frac{9\ln{(2/P_l)}}{\delta^2(4f^2-1)^2}.
\end{equation}
This means that the reduced fidelity in the projection states can be compensated by
increasing the number of experimental repetitions.

\section{Conclusion}
\label{sec:conclusion}
We have presented a measurement scheme by accessing the states of a quorum
for state tomography in the two-qubit space by realistic quantum gates
and three types of spin-to-charge conversion in a two-electron double-quantum dot.
The quantum gates include $\operatorname{SWAP}$ and $\sqrt{\operatorname{SWAP}}$ gates
realized by exchange interaction, as well as ESR rotations.
We assume the ESR gates to have the largest impact on the
fidelity of the quantum states.
Therefore, it is important that the ESR rotation is applied for
no more than a $\pi/2$ rotation per state.
Moreover, it should be an advantage that only one of the spins
has to be treated with ESR since in the experiment by Brunner and
coworkers,\cite{brunner} the fidelity of ESR was different for
the left and the right quantum dot.

Whether state tomography can be realized experimentally in the system
which we consider here depends on the realization of the
different quantum states.
A reduced fidelity in the experimental representation of the
(pure) quantum states does not prevent the tomographic
reconstruction in principle but reduces its quality.
The crucial point is the knowledge of the experimenter on which
(mixed) states the unknown quantum state is actually projected.
A lack of information about this state will in any case limit the precision
of the tomography.
Self-consistent measurements\cite{merkel,medford} or a calibration by analyzing
the tomography data\cite{shulman} can reduce this lack of knowledge.
In general, the accuracy of a realized tomography experiment
has to be determined experimentally as well.
Our measurement scheme is only one possible solution for
state tomography in the two-spin system. The question
which scheme is optimal regarding the accuracy of the estimated
density matrix with or without a limited number of experimental runs
remains open. The solution will crucially depend on the fidelity
of the different quantum gates performed in the system.

\begin{acknowledgments}
 We thank Cord A. M\"uller and Andr\'as P\'alyi for useful discussions and the DFG
for financial support under the programs SPP 1285
'Semiconductor spintronics' and SFB 767 'Controlled nanosystems'.
\end{acknowledgments}

\appendix

\section{Representation of a projection onto a state in matrix space}
\label{app:projector}
Assuming that we can perform any unitary operation in our four-level
system, we can project on any state
\begin{equation}
 \ket{\phi_j}= a_j \ket{\uparrow\uparrow} + b_j \ket{\uparrow\downarrow} + c_j \ket{\downarrow\uparrow} + d_j \ket{\downarrow\downarrow}
\end{equation}
where $|a_j|^2 + |b_j|^2 + |c_j|^2 + |d_j|^2 = 1$,
without loss of generality, $a_j \in \mathbbm{R}_{\ge0}$ as a global phase does not matter.\\
The corresponding projector can be written in the matrix form for the basis
$\{$
$\ket{\uparrow\uparrow}$,
$\ket{\uparrow\downarrow}$,
$\ket{\downarrow\uparrow}$,
$\ket{\downarrow\downarrow}$
$\}$,
\begin{equation}
 P_j = \ket{\phi_j}\bra{\phi_j} \equiv \left( \begin{array}{cccc} |a_j|^2  & a_jb_j^* & a_jc_j^* & a_jd_j^* \\
                                                                 b_ja_j^* & |b_j|^2  & b_jc_j^* & b_jd_j^* \\
                                                                 c_ja_j^* & c_jb_j^* & |c_j|^2  & c_jd_j^* \\
                                                                 d_ja_j^* & d_jb_j^* & d_jc_j^* & |d_j|^2
                                             \end{array}
\right) = \sum_{i,l=0,x,y,z} n_{jil}\frac{\sigma_{1i}\sigma_{2l}}{2}
\end{equation}
with

\begin{equation}
 \begin{split}
n_{j00}  = \frac{|a_j|^2 + |b_j|^2 + |c_j|^2 + |d_j|^2}{2}=\frac{1}{2}, & \hspace{1cm}
n_{j0x}  = \frac{a_j^*b_j+b_j^*a_j+c_j^*d_j+d_j^*c_j}{2},\\
n_{j0y}  = \frac{a_j^*b_j-b_j^*a_j+c_j^*d_j-d_j^*c_j}{2i}, & \hspace{1cm}
n_{j0z}  = \frac{|a_j|^2-|b_j|^2+|c_j|^2-|d_j|^2}{2},\\
n_{jx0}  = \frac{a_j^*c_j+b_j^*d_j+c_j^*a_j+d_j^*b_j}{2}, & \hspace{1cm}
n_{jxx}  = \frac{a_j^*d_j+d_j^*a_j+b_j^*c_j+c_j^*b_j}{2},\\
n_{jxy}  = \frac{a_j^*d_j-d_j^*a_j-b_j^*c_j+c_j^*b_j}{2i}, & \hspace{1cm}
n_{jxz}  = \frac{a_j^*c_j-b_j^*d_j+c_j^*a_j-d_j^*b_j}{2},\\
n_{jy0}  = \frac{a_j^*c_j+b_j^*d_j-c_j^*a_j-d_j^*b_j}{2i}, & \hspace{1cm}
n_{jyx}  = \frac{a_j^*d_j-d_j^*a_j+b_j^*c_j-c_j^*b_j}{2i},\\
n_{jyy}  = \frac{b_j^*c_j+c_j^*b_j-a_j^*d_j-d_j^*a_j}{2}, & \hspace{1cm}
n_{jyz}  = \frac{a_j^*c_j-b_j^*d_j-c_j^*a_j+d_j^*b_j}{2i},\\
n_{jz0}  = \frac{|a_j|^2 + |b_j|^2 - |c_j|^2 - |d_j|^2}{2}, & \hspace{1cm}
n_{jzx}  = \frac{a_j^*b_j+b_j^*a_j-c_j^*d_j-d_j^*c_j}{2},\\
n_{jzy}  = \frac{a_j^*b_j-b_j^*a_j-c_j^*d_j+d_j^*c_j}{2i}, & \hspace{1cm}
n_{jzz}  = \frac{|a_j|^2 - |b_j|^2 - |c_j|^2 + |d_j|^2}{2}.
\end{split}
\end{equation}

\section{No orthogonal basis in traceless matrix space by projectors on single pure states }
\label{app:proof}
For the reconstruction of the density matrix from the measurement data,
it would be ideal to have a set
$\{P_1,\ldots,P_{15}\}$ which fulfills for $i\neq j$
\begin{equation}
\label{eq:cond}
 \left<P_j-\frac{\mathbbm{1}}{4}\right|\!\left. P_i-\frac{\mathbbm{1}}{4}\right>_M = 0
~~\Leftrightarrow~~\braket{ P_j}{P_i}_M=\frac{1}{4}
~~\Leftrightarrow~~|\braket{\phi_j}{\phi_i}|^2=\frac{1}{4},
\end{equation}
which means that the traceless parts of the projection operators $P_j$
would be orthogonal with respect to
$\braket{\cdot}{\cdot}_M$.
Analogously, for a single spin, the density matrix can be determined
by projections on the $x$-, $y$-, and $z$-axis of the corresponding Bloch
sphere.
However, for our four-level system, Eq.~(\ref{eq:cond}) cannot be fulfilled
for 15 projection operators of the form  $P_j = \ket{\phi_j}\bra{\phi_j}$
which we show in this appendix.
For the proof we use the following lemma.

\textit{Lemma:} Let $V_1$ and $V_2$ be two real Euclidean spaces with dimension $n_1$ and $n_2$.
Let $\{a_1,\ldots,a_{n_1+n_2}\}$ be an orthogonal basis in the space $V_1\oplus V_2$ with
$|a_j|^2=c>0$, $|P_{V_1}a_j|^2=c_1$, and $|P_{V_2}a_j|^2=c_2$ $\forall j=1,\ldots,n_1+n_2$ where
$P_{V_1}$ ($P_{V_2}$) is the projection on the space $V_1$ ($V_2$).
Then $c_1$ and $c_2$ fulfill
\begin{equation}
 \frac{c_1}{c_2}=\frac{n_1}{n_2}.
\end{equation}
\textit{Proof of the Lemma:} 
We denote $a_1,\ldots,a_{n_1+n_2}$ in an orthonormal basis $\{e_1,\ldots,e_{n_1},e_{n_1+1},\ldots,e_{n_1+n_2}\}$
where the vectors $e_1,\ldots,e_{n_1}$ are completely in $V_1$ and
$e_{n_1+1},\ldots,e_{n_1+n_2}$ are completely in $V_2$.
Now we can write the normalized vectors $a_1/\sqrt{c},\ldots,a_{n_1+n_2}/\sqrt{c}$
as the columns of a matrix $A$ with the matrix elements
$A_{ij}=\braket{e_i}{a_j}/\sqrt{c}$.
The matrix $A$ is an orthogonal matrix, i.e., it fulfills
$A^TA=\mathbbm{1}$ but also $AA^T=\mathbbm{1}$, which reflects
the fact that rows and columns denote orthonormal bases in
$V_1\oplus V_2$.
From the conditions in the Lemma we find
\begin{equation}
 \sum_{i=1}^{n_1}A_{ij}^2=\frac{c_1}{c}~~\text{and}~~\sum_{i=n_1+1}^{n_1+n_2}A_{ij}^2=\frac{c_2}{c},
\end{equation}
and thus
\begin{equation}
\label{eqn:cs}
 \sum_{j=1}^{n_1+n_2}\sum_{i=1}^{n_1}A_{ij}^2=(n_1+n_2)\frac{c_1}{c}~~\text{and}~~\sum_{j=1}^{n_1+n_2}\sum_{i=n_1+1}^{n_1+n_2}A_{ij}^2=(n_1+n_2)\frac{c_2}{c}.
\end{equation}
On the other hand, using that the rows of $A$ denote orthonormal vectors in $\mathbbm{R}^{n_1+n_2}$
yields
\begin{equation}
\label{eqn:ns}
 \sum_{i=1}^{n_1}\sum_{j=1}^{n_1+n_2}A_{ij}^2=\sum_{i=1}^{n_1} 1 = n_1~~\text{and}~~\sum_{i=n_1+1}^{n_1+n_2}\sum_{j=1}^{n_1+n_2}A_{ij}^2=\sum_{i=n_1+1}^{n_1+n_2} 1 = n_2.
\end{equation}
Building the ratio directly leads to
\begin{equation}
 \frac{\sum_{j=1}^{n_1+n_2}\sum_{i=1}^{n_1}A_{ij}^2}{\sum_{j=1}^{n_1+n_2}\sum_{i=n_1+1}^{n_1+n_2}A_{ij}^2} \stackrel{(\ref{eqn:cs})}=  \frac{c_1}{c_2} \stackrel{(\ref{eqn:ns})}= \frac{n_1}{n_2}.\blacksquare
\end{equation}

Now we come back to the 15 basis states for the space of traceless
Hermitian matrices constructed by pure states in four-dimensional Hilbert space.
We use the notation from Appendix \ref{app:projector}.
Assume that $\ket{\phi_1}=\ket{\uparrow\uparrow}$, if it is not,
we can perform a unitary transformation to match this starting point.
The corresponding projector is
$P_1=(\mathbbm{1}+\sigma_{1z}+\sigma_{2z}+\sigma_{1z}\sigma_{2z})/4$.
It follows that $a_j=1/2$ in order to fulfill
$|\braket{\phi_1}{\phi_j}|^2=1/4$ for $j=2,\ldots,15$.
Therefore we get
\begin{equation}
\label{eqn:b2}
 \begin{split}
n_{j00}  =\frac{1}{2}, & \hspace{1cm}
n_{j0x}  = \frac{\operatorname{Re} b_j+c_j^*d_j+d_j^*c_j}{2},\\
n_{j0y}  = \frac{\operatorname{Im} b_j+c_j^*d_j-d_j^*c_j}{2i}, & \hspace{1cm}
n_{j0z}  = \frac{\frac{1}{4}-|b_j|^2+|c_j|^2-|d_j|^2}{2},\\
n_{jx0}  = \frac{\operatorname{Re} c_j+b_j^*d_j+d_j^*b_j}{2}, & \hspace{1cm}
n_{jxx}  = \frac{\operatorname{Re} d_j+b_j^*c_j+c_j^*b_j}{2},\\
n_{jxy}  = \frac{\operatorname{Im} d_j-b_j^*c_j+c_j^*b_j}{2i}, & \hspace{1cm}
n_{jxz}  = \frac{\operatorname{Re} c_j-b_j^*d_j-d_j^*b_j}{2},\\
n_{jy0}  = \frac{\operatorname{Im} c_j+b_j^*d_j-d_j^*b_j}{2i}, & \hspace{1cm}
n_{jyx}  = \frac{\operatorname{Im} d_j+b_j^*c_j-c_j^*b_j}{2i},\\
n_{jyy}  = \frac{b_j^*c_j+c_j^*b_j-\operatorname{Re} d_j}{2}, & \hspace{1cm}
n_{jyz}  = \frac{\operatorname{Im} c_j-b_j^*d_j+d_j^*b_j}{2i},\\
n_{jz0}  = \frac{\frac{1}{4} + |b_j|^2 - |c_j|^2 - |d_j|^2}{2}, & \hspace{1cm}
n_{jzx}  = \frac{\operatorname{Re} b_j-c_j^*d_j-d_j^*c_j}{2},\\
n_{jzy}  = \frac{\operatorname{Im} b_j-c_j^*d_j+d_j^*c_j}{2i}, & \hspace{1cm}
n_{jzz}  = \frac{\frac{1}{4} - |b_j|^2 - |c_j|^2 + |d_j|^2}{2}.
\end{split}
\end{equation}
Now we define a new basis of the space of $4\times 4$ matrices,
$\{\tau_i; i=0,\ldots, 15\}$ with
\begin{equation}
 \begin{split}
  \tau_0=\frac{\mathbbm{1}}{2},\hspace{1cm} &
  \tau_1=\frac{\sigma_{2z}+\sigma_{1z}+\sigma_{1z}\sigma_{2z}}{2\sqrt{3}},\\
  \tau_2=\frac{\sigma_{1x}+\sigma_{1x}\sigma_{2z}}{2\sqrt{2}},\hspace{1cm}&
  \tau_3=\frac{\sigma_{2x}+\sigma_{1z}\sigma_{2x}}{2\sqrt{2}},\\
  \tau_4=\frac{\sigma_{1y}+\sigma_{1y}\sigma_{2z}}{2\sqrt{2}},\hspace{1cm}&
  \tau_5=\frac{\sigma_{2y}+\sigma_{1z}\sigma_{2y}}{2\sqrt{2}},\\
  \tau_6=\frac{\sigma_{1x}\sigma_{2x}-\sigma_{1y}\sigma_{2y}}{2\sqrt{2}},\hspace{1cm}&
  \tau_7=\frac{\sigma_{1x}\sigma_{2y}+\sigma_{1y}\sigma_{2x}}{2\sqrt{2}},\\
  \tau_8=\frac{\sigma_{1x}-\sigma_{1x}\sigma_{2z}}{2\sqrt{2}},\hspace{1cm}&
  \tau_9=\frac{\sigma_{2x}-\sigma_{1z}\sigma_{2x}}{2\sqrt{2}},\\
  \tau_{10}=\frac{\sigma_{1y}-\sigma_{1y}\sigma_{2z}}{2\sqrt{2}},\hspace{1cm}&
  \tau_{11}=\frac{\sigma_{2y}-\sigma_{1z}\sigma_{2y}}{2\sqrt{2}},\\
  \tau_{12}=\frac{\sigma_{1x}\sigma_{2x}+\sigma_{1y}\sigma_{2y}}{2\sqrt{2}},\hspace{1cm}&
  \tau_{13}=\frac{\sigma_{1x}\sigma_{2y}-\sigma_{1y}\sigma_{2x}}{2\sqrt{2}},\\
  \tau_{14}=\frac{\sigma_{1z}-\sigma_{2z}}{2\sqrt{2}},\hspace{1cm}&
  \tau_{15}=\frac{\sigma_{1z}+\sigma_{2z}-2\sigma_{1z}\sigma_{2z}}{2\sqrt{6}},
 \end{split}
\end{equation}
which again fulfills $\braket{\tau_k}{\tau_l}_M=\delta_{kl}$.
$P_2-\mathbbm{1}/4,\ldots,P_{15}-\mathbbm{1}/4$ are now supposed to be
an orthogonal basis in the space spanned by $\tau_2,\ldots,\tau_{15}$.
We show that this is not possible for the following reason.
When we expand $P_j$ in the new basis, $P_j = \sum_i m_{ji}\tau_i$,
we find easily for $j\ge2$ from Eq.~(\ref{eqn:b2}) that
\begin{equation}
\begin{split}
 &m_{j0}=\frac{1}{2},~~m_{j1}=0,\\
 &m_{j2}=\frac{\operatorname{Re} c_j}{\sqrt{2}},~~m_{j3}=\frac{\operatorname{Re} b_j}{\sqrt{2}},~~m_{j4}=\frac{\operatorname{Im} c_j}{\sqrt{2}}\\
 &m_{j5}=\frac{\operatorname{Im} b_j}{\sqrt{2}},~~m_{j6}=\frac{\operatorname{Re} d_j}{\sqrt{2}},~~m_{j7}=\frac{\operatorname{Im} d_j}{\sqrt{2}}
\end{split}
\end{equation}
and thus
\begin{equation}
\label{eq:norm1}
 \sum_{i=2}^7m_{ji}^2 = \frac{|b_j|^2+|c_j|^2+|d_j|^2}{2}=\frac{1-|a_j|^2}{2}=\frac{3}{8}.
\end{equation}
On the other hand due to $\operatorname{tr}(P_j) = 1$ and $P_j^2=P_j$, we find for $j=2,\ldots,15$
\begin{equation}
 1=\braket{ P_j}{P_j}_M = \sum_{i=0}^{15} m_{ji}^2=m_{j0}^2+m_{j1}^2+\sum_{i=2}^{15}m_{ji}^2=\frac{1}{4}+\sum_{i=2}^{15} m_{ji}^2,
\end{equation}
which yields
\begin{equation}
 \sum_{i=2}^{15} m_{ji}^2 = \frac{3}{4}.
\end{equation}
Using Eq.~(\ref{eq:norm1}) we find
\begin{equation}
 \sum_{i=8}^{15}m_{ji}^2 = \frac{3}{4}-\sum_{i=2}^7m_{ji}^2=\frac{3}{8}
\end{equation}
and thus 
\begin{equation}
 \frac{\sum_{i=2}^7m_{ji}^2}{\sum_{i=8}^{15}m_{ji}^2}=1\neq\frac{6}{8}
\end{equation}
which means that the ratio of the projections of $P_j$ on the subspace
$\operatorname{span} \{\tau_2,\ldots,\tau_7\}$ and $\operatorname{span}\{\tau_8,\ldots,\tau_{15}\}$
is a fixed value which does not correspond to the dimensions of these
subspaces and thus violates the lemma above.
Therefore an orthogonal basis for $\operatorname{span}\{\tau_1,\ldots,\tau_{15}\}$
cannot be constructed by $P_1-\mathbbm{1}/4,\ldots,P_{15}-\mathbbm{1}/4$. $\blacksquare$

\section{Reconstruction with mutually unbiased bases}
\label{app:mub}
Here we show that $|\det(\mathcal{P})|=\frac{1}{32}$ if the matrix $\mathcal{P}$
as introduced in Sec.~\ref{sec:general} is gained from five mutually unbiased
bases, including three states from each basis,
$\{\ket{\phi_{01}},\ket{\phi_{02}},\ket{\phi_{03}},\ket{\phi_{11}},\ldots,\ket{\phi_{43}}\}$.
The rows of $\mathcal{P}$ represents the traceless parts of the
corresponding projection operators
$P_{3j+k}=\ket{\phi_{jk}}\bra{\phi_{jk}}$, which we denote
\begin{equation}
\mathcal{P}_{l}=\left(\braket{P_l}{D_1}_M,\ldots,\braket{P_l}{D_{15}}_M\right).
\end{equation}
We want to apply Gram-Schmidt orthogonalization on the row vectors $\mathcal{P}_{l}$
which leaves the determinant of $\mathcal{P}$ invariant.
As
\begin{equation}
\mathcal{P}_{3j+k}\mathcal{P}_{3i+l}^T=|\braket{\phi_{jk}}{\phi_{il}}|^2-\frac{1}{4}=\delta_{ij}\left(\delta_{kl}-\frac{1}{4}\right)\text{ with }i,j\in\{1,\ldots,5\}\text{ and }k,l\in\{1,2,3\}
\end{equation}
we only have to orthogonalize the five three-dimensional subspaces for $i=j$.
From $\mathcal{P}_{3j+k}\mathcal{P}_{3j+l}^T=\delta_{kl}-\frac{1}{4}$ we gain easily the
orthogonalized row vectors from the Gram-Schmidt method,
\begin{equation}
 \begin{split}
  \mathcal{P'}_{3j+1} & = \mathcal{P}_{3j+1},\\
  \mathcal{P'}_{3j+2} & = \mathcal{P}_{3j+2} + \frac{1}{3}\mathcal{P}_{3j+1},\\
  \mathcal{P'}_{3j+1} & = \mathcal{P}_{3j+3} + \frac{1}{2}\mathcal{P'}_{3j+2} + \frac{1}{3}\mathcal{P}_{3j+1},
 \end{split}
\end{equation}
which have the lengths $\sqrt{\frac{3}{4}}$, $\sqrt{\frac{2}{3}}$,
and $\sqrt{\frac{1}{2}}$.
A unitary transformation does not change the determinant
and transforms $\mathcal{P'}$ to a diagonal matrix.
Therefore, we find
\begin{equation}
 \det(\mathcal{P}) = \det(\mathcal{P'})
= \left(\sqrt{\frac{3}{4}}\cdot\sqrt{\frac{2}{3}}\cdot\sqrt{\frac{1}{2}}\right)^5
= \left(\frac{1}{2}\right)^5
= \frac{1}{32}.
\end{equation}

\end{document}